\documentclass[letterpaper]{article} % DO NOT CHANGE THIS
\usepackage{aaai23}  % DO NOT CHANGE THIS
\usepackage{times}  % DO NOT CHANGE THIS
\usepackage{helvet}  % DO NOT CHANGE THIS
\usepackage{courier}  % DO NOT CHANGE THIS
\usepackage[hyphens]{url}  % DO NOT CHANGE THIS
\usepackage{graphicx} % DO NOT CHANGE THIS
\usepackage{natbib}  % DO NOT CHANGE THIS AND DO NOT ADD ANY OPTIONS TO IT
\usepackage{caption} % DO NOT CHANGE THIS AND DO NOT ADD ANY OPTIONS TO IT
\usepackage{appendix}
\usepackage{algorithm}
\usepackage{algorithmic}
\def\Mat#1{{\boldsymbol{#1}}}
\def\Vec#1{{\boldsymbol{#1}}}
\def\eg{\emph{e.g}}
\def\ie{\emph{i.e}}

\usepackage{newfloat}
\usepackage{listings}
\usepackage{graphicx}
\usepackage{multirow}
\usepackage{booktabs}
\usepackage{amsmath}
\usepackage{amssymb}
\usepackage{cleveref}
\DeclareCaptionStyle{ruled}{labelfont=normalfont,labelsep=colon,strut=off} % DO NOT CHANGE THIS
\floatstyle{ruled}
\newfloat{listing}{tb}{lst}{}
\floatname{listing}{Listing}
\title{\emph{AHEAD}: A Triple Attention Based Heterogeneous Graph \\ Anomaly Detection Approach}
\author {
    Shujie Yang\textsuperscript{\rm 1},
    Binchi Zhang\textsuperscript{\rm 2},
    Shangbin Feng\textsuperscript{\rm 3},
    Zhanxuan Tan\textsuperscript{\rm 1},\\
    Qinghua Zheng\textsuperscript{\rm 1},
    Jun Zhou\textsuperscript{\rm 4},
    Minnan Luo\textsuperscript{\rm 1}
}
\affiliations {
    \textsuperscript{\rm 1}Xi’an Jiaotong University, Xi’an, China
    \textsuperscript{\rm 2}University of Virginia, Charlottesville, USA\\
    \textsuperscript{\rm 3}University of Washington, Seattle, USA
    \textsuperscript{\rm 4}Ant Financial, Hangzhou, China\\
    yangshujie@stu.xjtu.edu.cn, epb6gw@virginia.edu, shangbin@cs.washington.edu, tanzhaoxuan@stu.xjtu.edu.cn, qhzheng@mail.xjtu.edu.cn, jun.zhoujun@antfin.com, minnluo@xjtu.edu.cn
}

\begin{document}

\maketitle

\begin{abstract}
Graph anomaly detection on attributed networks has become a prevalent research topic due to its broad applications in many influential domains. In real-world scenarios, nodes and edges in attributed networks usually display distinct heterogeneity, \ie. attributes of different types of nodes show great variety, different types of relations represent diverse meanings.
Anomalies usually perform differently from the majority in various perspectives of heterogeneity in these networks.
However, existing graph anomaly detection approaches do not leverage heterogeneity in attributed networks, which is highly related to anomaly detection. In light of this problem, we propose AHEAD: a heterogeneity-aware unsupervised graph anomaly detection approach based on the encoder-decoder framework. Specifically, for the encoder, we design three levels of attention, \ie. attribute level, node type level, and edge level attentions to capture the heterogeneity of network structure, node properties and information of a single node, respectively. In the decoder, we exploit structure, attribute, and node type reconstruction terms to obtain an anomaly score for each node. Extensive experiments show the superiority of \emph{AHEAD} on several real-world heterogeneous information networks compared with the state-of-arts in the unsupervised setting. Further experiments verify the effectiveness and robustness of our triple attention, model backbone, and decoder in general. 
\end{abstract}

\begin{figure}[t]
    \centering
    \includegraphics[width=0.9\linewidth]{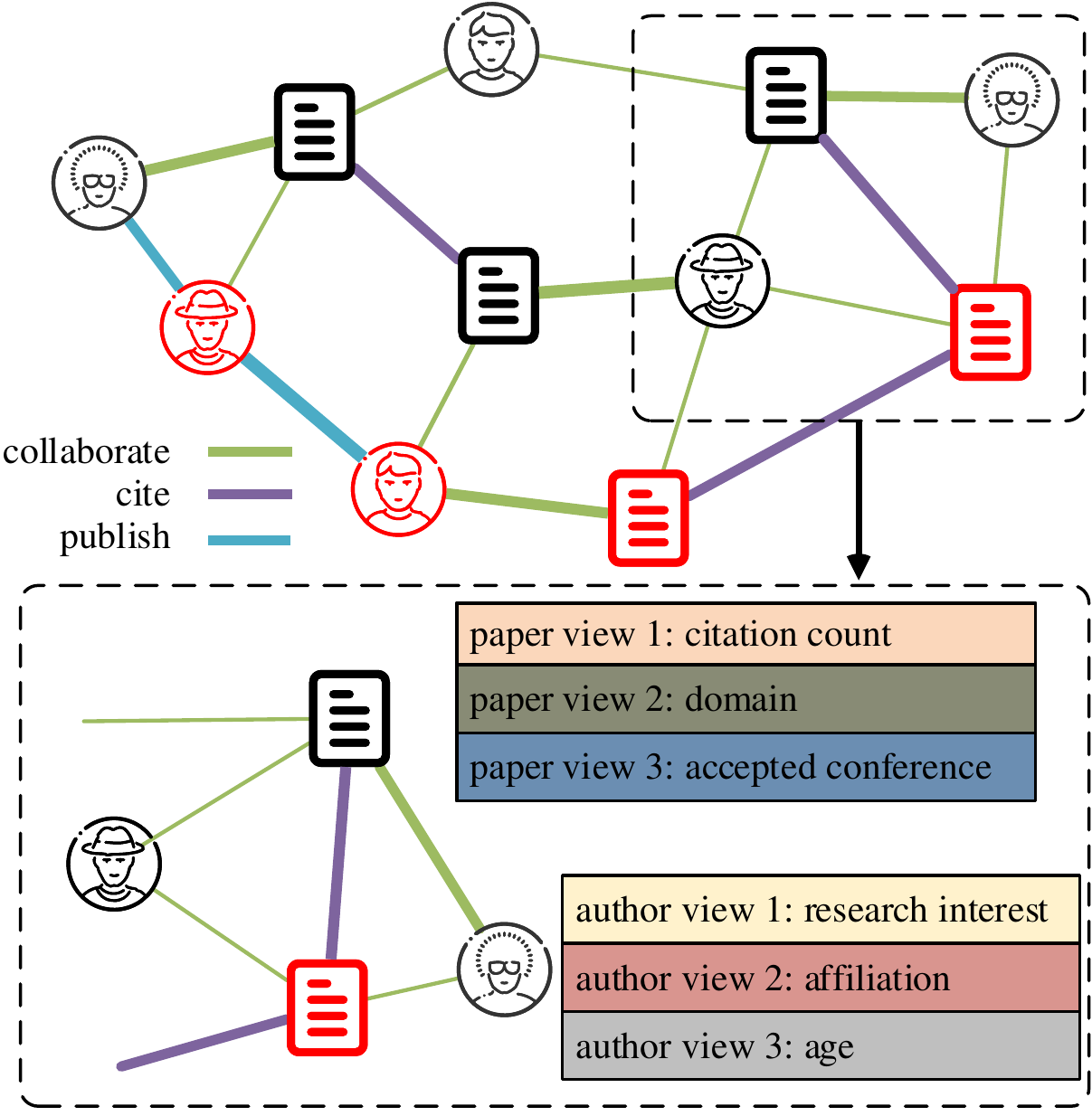}
    \caption{An example to illustrate that anomalies can be associated with node, edge heterogeneity ($\boldsymbol{above}$), and attribute characteristics heterogeneity ($\boldsymbol{below}$). }
    \label{fig:teaser}
\end{figure}

\section{Introduction}
Graph anomaly detection has become an important research topic for its broad applications in many high-impact areas, \eg. spammer detection in social media \citep{akoglu2015graph, fraud}, fraud detection \citep{huang2018codetect}, network intrusion detection \citep{lo2021graphsage}, and link prediction in social network \cite{social1}. Typically, graph anomaly detection aims at recognizing deviant samples and unusual patterns from attributed networks. In attributed networks, unusual patterns often exist in both graph structure and node features. Therefore, how to effectively leverage the information of structure and feature simultaneously is the crucial problem in graph anomaly detection.

Towards this problem, many researchers investigated different graph anomaly detection methods. Different from normal classification tasks, most graph anomaly detection tasks have no ground truth because the labeling process is extremely difficult. It has been shown that humans can only perform at best as good as random in labeling a review as fake or not based on its text \citep{ott2011finding,akoglu2015graph}. Therefore, existing methods for graph anomaly detection are usually unsupervised. \citet{CODA} proposed a probabilistic graph anomaly detection model based on hidden Markov random field, where they exploit a generative mixture model to describe node feature information and spatial constraints to describe graph structure information. \citet{Radar} proposed a graph anomaly detection framework that calculates anomaly score based on residual of original and reconstructed feature matrix and graph structure. The linear map in the reconstruction step makes the model more powerful. \citet{DOMINANT} proposed an encoder-decoder framework where the encoder is based on graph convolutional networks \citep{GCN}, and the decoder is a reconstruction function. The deep neural network further improves the residual model. \citet{ALARM} introduced multi-view split into node features to better leverage the feature information of different semantics in the encoder-decoder framework.

Although existing approaches have made great success in graph anomaly detection tasks, none of them consider the heterogeneity in real-world attributed networks. Real-world attributed networks are usually formed as heterogeneous information networks (HINs), which contain different types of nodes and relations to incorporate more information and contain richer semantics in nodes and links. The heterogeneity of HINs exists universally in attributes, nodes, and edges. As Fig. \ref{fig:teaser} shows, deviant patterns in HINs could deeply concern the heterogeneity. This problem requires anomaly detection frameworks to discriminate different types of information at the attribute level, the node level, and the edge level. However, in most existing graph anomaly detection approaches, all the nodes and edges are evenly treated. Different types of information in a node feature are embedded into the same space, which means the heterogeneity information is not fully leveraged in existing methods.

In light of this problem, we propose \emph{AHEAD}, a novel heterogeneity-aware unsupervised graph anomaly detection framework based on encoder-decoder architecture. To capture the heterogeneity, we embed different types of nodes into different embedding spaces and exploit triple attentions, \ie. attribute level, node type level, and edge level, to obtain the importance of various heterogeneous information. Specifically, in the encoder step, we introduce multi-view split into the feature space of different node types. We exploit heterogeneous graph transformer \citep{HGT} on graphs consisting of single views in all cases of view combination where anomalies might exist. Consequently, we exploit the embeddings of different views to reconstruct the node type matrix, feature matrices of different node types, and adjacency matrices of different relations. Then we obtain the anomaly score by calculating the reconstruction error with corresponding attention weights. Our contributions are summarized as follows:
\begin{itemize}
    \item We propose a heterogeneity-aware graph anomaly detection framework which could better leverage heterogeneous information in real-world graph-structured data.
    \item We introduce triple attention to capture the importance of different relations, node types, and information of a single node.
    \item Extensive experiments on four real-world datasets demonstrate that \emph{AHEAD} outperforms all existing baselines, which verifies the effectiveness of \emph{AHEAD} on heterogeneous graph anomaly detection compared with state-of-the-arts.
\end{itemize}

\begin{figure*}[t]
    \centering{\includegraphics[width=\textwidth]{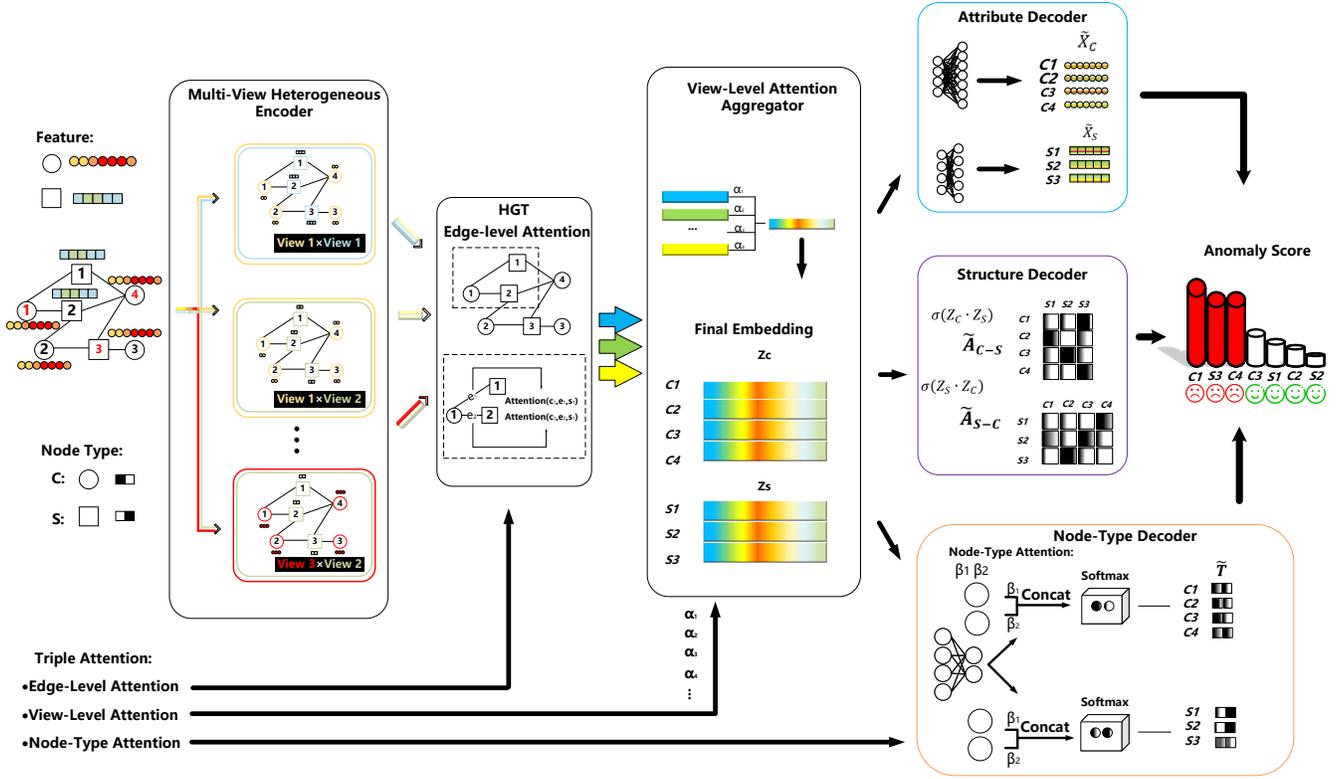}}
    \caption{The proposed triple attention based heterogeneous graph anomaly detection model \emph{AHEAD}. Triple attention refers to edge-level attention, view-level attention and node-type attention. There are two types of nodes: $C, S$ in the given heterogeneous graph and $C$-types nodes have three views and $S$-types nodes have two views. Every combination of views is input into a HGT separately and generates an overall embedding through aggregation based on view-level attention. After decoded by structure \& attribute \& node-type decoders, anomalous can be detected by ranking anomaly scores.}
    \label{overview}
\end{figure*}

\section{RELATED WORK}
\subsection{Anomaly Detection On Graph}
Anomaly detection refers to the process of detecting instances that deviate significantly from most instances. 
%As a popular research area, many methods have been introduced to implement anomaly detection. 
There have been many unsupervised anomaly detection methods on graphs \cite{gutierrez2020multi, zhang2022reconstruction, qiu2022raising, ding2021inductive, luo2022comga, ma2022deep, zhao2021action, cai2021structural, zhou2021subtractive, jin2021anemone, sun2020anomaly} and most of them are based on autoencoders. \emph{DOMINANT} \cite{DOMINANT} feeds structure information and all node attribute information of graph to GCN to obtain the embedding of nodes, and reconstructs the original data through an autoencoder to detect anomaly nodes. \emph{ALARM} \cite{ALARM} consists of a multi-view encoder, an aggregator, and a structure \& attribute decoder.%\emph{GCNAE} \citep{GCNAE} proposes graph automatic encoder ($GAE$) and variational graph automatic encoder ($VGAE$) based on automatic encoder and variational automatic encoder. $GAE$ is used to obtain appropriate embedding to represent the nodes in the graph. $VGAE$ can obtain the embedding of nodes in the graph through the structure of encoder-decoder. \emph{DONE} \citep{DONE} consists of an attribute autoencoder and a structure autoencoder. By optimizing reconstruction loss can anomalies be detected. Such methods usually consist of an encoder and a decoder. Encoder compresses the original feature into low-dimensional space while the decoder reconstructs them back to the original dimension. 
%These methods combine detection with anomaly scoring in some ways and are based on the assumption that normal instances can be better restructured from compressed space than anomalies, and have excellent experimental detection performance.
%However, many challenges about anomaly detection remain to be tackled. For example, the diversity and heterogeneity of anomalies are still difficult to be learned. Although there have been many fraud or user detection methods on heterogeneous graphs, anomaly detection is different from them in many aspects like applicable scenarios and characteristics of the target.
These methods calculate anomaly scores based on the assumption that normal instances can be better restructured from hidden space than anomalies and regard nodes with high anomaly scores as anomalies. Although they have satisfying performance, the diversity and heterogeneity of anomalies are still difficult to be learned. Although there have been many fraud or user detection methods on heterogeneous graphs \cite{li2021live, hei2021hawk, zhong2020financial, li2021relevance, huang2020heterogeneous, yuan2019jointly, liu2018heterogeneous}, anomaly detection is different from them in many aspects like applicable scenarios and characteristics of the target. For instance, in \emph{IHGAT} \cite{liu2021intention}, a heterogeneous network is devised to detect fraud transactions by leveraging the information over two types of nodes and edges in the network. This method can only be applied in specific networks, which consist of two types of nodes: transaction and intention nodes, and two types of edges: transaction-intention and transaction-transaction edges, and can only detect specific type of anomaly: fraud transactions. However, \emph{AHEAD} can be applied in any heterogeneous network and can recognize any anomaly, even if it is not defined explicitly.
\subsection{Heterogeneous Graph Learning}
At present, most graph learning methods are based on homogeneous graphs \cite{chen2019deep}. 
% The representation learning of heterogeneous graphs faces several challenges: how to solve the heterogeneity of nodes and edges, how to embody the heterogeneity in the final embedding of graphs, etc. 
Heterogeneous graph refers to a graph with more than one type of nodes or edges. The early studies of heterogeneous graph learning are mostly based on meta-path \cite{sun2011pathsim} \cite{sun2012mining}, which is a specific path connecting two entities. For example, \emph{HAN} \cite{HAN} proposes a new attention mechanism of heterogeneous graphs, which involves node-level and semantic-level. Node-level attention mainly learns the weights between nodes and their adjacent nodes. Semantic-level attention learns the weights based on different meta-paths. \emph{MAGNN} \cite{IMDB} proposes node content transformation, intra-metapath aggregation, and inter-metapath aggregation to embed a heterogeneous graph. 
%Besides, other meta-path-based methods \cite{sun2011pathsim, dong2017metapath2vec} make much progress.

However, meta-path-based methods have some limitations \cite{IMDB}. Meta-path doesn't leverage node features and only considers two end nodes ignoring all the intermediate nodes along the meta-path, which contributes to information loss. What's more, meta-path needs to be manually designed, which is complicated. Therefore, novel methods eliminating meta-path emerge. \emph{HGT} \cite{HGT} proposes an attention mechanism of each edge, which makes it possible to generate dedicated representations for different types of nodes and edges. Besides, other non meta-path-based methods \cite{rgt, rshn, HGNNAC, Het, HetSANN, simplehgn} are effective, too. In this work, we adopt \emph{HGT} \cite{HGT} as the backbone to explore how to integrate heterogeneity into graph anomaly detection.
% \emph{HGNN-AC} \cite{HGNNAC} proposed a method to deal with non-attribute nodes in heterogeneous graphs, which completes the attributes of non-attribute nodes by weighted aggregation of the attributed nodes under the guidance of the topological relationship learned by other methods. \emph{HetGNN} \cite{Het} samples heterogeneous neighbors by random walk, and then aggregates attribute and node-type information from them. 

\subsection{Multi-View Representation Learning}
%Multi-view representation learning is based on multi-view data, which can be described as a multi measurement of basic information. It acquires the representation containing useful information and employs it in downstream machine learning tasks \cite{li2013stochastic, han2021multi}. Generally, data from different views contain complementary information, which makes multi-view representation learning acquire more knowledge than normal ways. \emph{CCA} \citep{CCA} with its kernel extensions \citep{Kernel} tries to find two linear projections for corresponding views to maximize correlations between the transformed variables. However, \emph{CCA} can not capture high-level correlation on multi-view data. In order to solve this defect, traditional methods and network-based methods develop constantly. For instance, \citet{2015Multi} propose a multi-view CNN, which integrates information from multiple 2D views into a single representation based on convolutional neural networks. 
Multi-view representation learning is based on multi-view data, which can be described as a multi-measurement of basic information, to acquire the representation containing useful information and employ it in downstream machine learning tasks \cite{li2013stochastic, han2021multi}. Generally, data from different views contain complementary information, which makes multi-view representation learning acquire more knowledge than the single view. CCA \citep{CCA} with its kernel extensions \citep{Kernel} is regarded as an early pioneering study of multi-view representation learning, which tries to find two linear projections for corresponding views to maximize correlations between the transformed variables. In order to solve the defect that CCA can not capture high-level correlation on multi-view data, traditional methods and network-based methods develop constantly. A large number of variants are proposed. What they have in common is that they explore the complementary information of multi-view to comprehensively represent the data. For instance, \citet{2015Multi} propose a multi-view CNN, which integrates information from multiple 2D views into a single representation based on convolutional neural networks. 

\section{Methodology}
\subsection{Preliminary}
Given a heterogeneous graph $\mathcal{G}=\left(\mathcal{V},\mathcal{E},\mathcal{A},\mathcal{R}\right)$, where sets $\mathcal{V}$ and $\mathcal{E}$ collect the nodes and edges, respectively. Each node $v\in\mathcal{V}$ and edge $e\in\mathcal{E}$ are associated with their type mapping functions $\tau:\mathcal{V}\rightarrow\mathcal{A}$ and $\phi:\mathcal{E}\rightarrow\mathcal{R}$ respectively. Without loss of generality, we assume that each node $v\in\mathcal{V}$ is associated with a node attribute vector $\Vec{x}_v$. Attributes associated with same type of nodes have the same dimension. For nodes with node type $a\in\mathcal{A}$, there are $k_a$ views in their attribute matrix. It can be divided as $\Mat{X}_a=\left[\Mat{X}_a^{\left(1\right)},  \Mat{X}_a^{\left(2\right)}, \cdots\ \Mat{X}_a^{\left(k_a\right)}\right]$, according to these $k_a$ views. We have $\Mat{X}_a^{\left(i\right)}\in\mathbb{R}^{n_a\times D_a^{\left(i\right)}}$, where $n_a$ refers to the count of nodes with type $a$, $D_a^{\left(i\right)}$ represents the dimension of the $i$-th view. To be more specific, for node $v\in\mathcal{V}$ and $\tau(v)\in\mathcal{A}$, its attribute vector $\Mat{x}_{\tau(v)}$ is $\Mat{x}_{\tau(v)}=\left[\Mat{x}_{\tau(v)}^{\left(1\right)}\, \Mat{x}_{\tau(v)}^{\left(2\right)}\, \cdots\ \Mat{x}_{\tau(v)}^{\left(k_{\tau(v)}\right)}\right]$, where $\Mat{x}_{\tau(v)}^{\left(i\right)}\in\mathbb{R}^{D_{\tau(v)}^{\left(i\right)}}$ for $i=1,2,\cdots,k_{\tau(v)}$.

Adjacency matrix describes the topology structure of a graph $\mathcal{G}$. However, in a heterogeneous graph, adjacency matrices are related to edge types. Given a target node $t$ and all $t$'s neighbours $s\in N(t)$, tuples $(s,e,t)$ denote a relation of two nodes $s,t\in\mathcal{V}$ and an edge $e\in\mathcal{E}$ connecting them and triplets $<\tau(s),\phi(e),\tau(t)>$ denote the meta realtion. For every $(s,e,t)$, adjacency matrix is $\Mat{A}_{\phi(e)}\in\mathbb{R}^{n_{\tau(s)}\times n_{\tau(t)}}$, where $n_{\tau(t)}$ is the number of $\tau(t)$-type nodes. Note that adjacency matrix is indexed by edge type, which means every edge type has a unique adjacency matrix.
%It is worth noting that, in undirected graphs, we can easily consider $\Mat{A}_{\phi(e^{-1})}=\Mat{A}_{\phi(e)}^T$. However, in directed graphs, $\Mat{A}_{\phi(e^{-1})}$ and $\Mat{A}_{\phi(e)}^T$
%are usually not the same thing, it's necessary to consider separately.

%\subsection{Overview}
In this section, we propose \emph{AHEAD}, an unsupervised multi-view deep learning model based on triple attention: edge-level attention, view-level attention, and node-type attention. Fig. \ref{overview} illustrates the architecture schematically. The entire model is composed of an multi-view heterogeneous graph encoder, a view-level weighed aggregator, and a structure \& attribute \& node-type decoder.% (1) Multi-view heterogeneous graph encoder: using multi-view heterogeneous graph transformer based on edge-level attention to embed given heterogeneous graphs. (2) view-level attention aggregator: aggregating every embedding based on view-level attention. (3) structure\&attribute\&node-type decoder: reconstructing graph structure, node attributes and node-type features. We propose that the greater the difference between the reconstructed features and the original features, the instance is more likely to be anomalous. So we can rank anomaly probability by calculating structure \& attribute \& node-type reconstruction errors.
\subsection{Multi-View Heterogeneous Graph Encoder}
In heterogeneous graphs, nodes and edges are various. When obtaining the embedding of a node by aggregating the attributes of neighbour nodes, naturally we hope the network can recognize the variety of edges and learn the importance of each edge, so that embedding of nodes can be better obtained in aggregation based on importance. Networks like GCN 
could not recognize edge heterogeneity. HGT \cite{HGT}, in which the network uses the attention mechanism to automatically learn the importance of each edge, provides us with tools. Hence we will exploit HGT as the encoder to encode given heterogeneous graphs.% The core idea of HGT is to parameterize the weight matrix of heterogeneous mutual attention, message transmission, and propagation steps by using the meta relationship of heterogeneous graphs.

In order to obtain the embedding of target node $t$ from source node $s$ in $(s,e,t)$, heterogeneous multi-head attention $\boldsymbol{Attention}_{edge}(s,e,t)$ denoted by $\boldsymbol{Att}_{edge}(s,e,t)$ and message $\boldsymbol{Message}(s,e,t)$ denoted by $\boldsymbol{Mes}(s,e,t)$ should be computed by \cite{HGT}
\begin{equation*}
    \widetilde{\Mat{H}}^{(l)}[t]=\underset{\forall s\in N\left(t\right)}{\sum}\boldsymbol{Att}_{edge}(s,e,t)\times\boldsymbol{Mes}(s,e,t)
\end{equation*}

\begin{equation*}
\boldsymbol{Att}_{edge}(s,e,t)=
\underset{\forall s\in N(t)}{Softmax}\left(\underset{i\in\left[1,h\right]}{\boldsymbol{Concat}}\  Att-head^i_{(s,e,t)}\right)
\end{equation*}

\begin{equation*}
    \boldsymbol{Mes}(s,e,t)=\underset{i\in\left[1,h\right]}{\boldsymbol{Concat}}\ Mes-head^i_{(s,e,t)}
\end{equation*}
% \begin{equation}
% \begin{aligned}
%     \boldsymbol{Attention}_{edge}(s,e,t)&=\underset{\forall s\in N(t)}{Softmax}\left(\underset{i\in\left[1,h\right]}{\boldsymbol{Concat}}\  Att-head^i_{(s,e,t)}\right)\\
%     Att-head^i_{(s,e,t)}&=K^i\left(s\right)W^{Att}_{\phi(e)}Q^i\left(t\right)^T\frac{\mu_{<\tau(s),\phi(e),\tau(t)>}}{\sqrt{d}}\\
%     K^i\left(s\right)&=K-Linear_{\tau(s)}^i\left(H^{(l-1)}[s]\right)\\
%     Q^i\left(t\right)&=Q-Linear_{\tau(t)}^i\left(H^{(l-1)}[t]\right)\\
% \end{aligned}
% \end{equation}
% \begin{equation}
%     \begin{aligned}
%     &\boldsymbol{Message}(s,e,t)=\underset{i\in\left[1,h\right]}{\boldsymbol{Concat}}\ Mes-head^i_{(s,e,t)}\\
%     &Mes-head^i_{(s,e,t)}=M-Linear^i_{\tau(s)}\left(H^{(l-1)}[s]\right)W^{Mes}_{\phi(e)}\\
%     \end{aligned}
% \end{equation}
where the $i$-th attention head is calculated by
\begin{align}
    Att-head^i_{(s,e,t)}=&K^i\left(s\right)W^{Att}_{\phi(e)}Q^i\left(t\right)^T\frac{\mu_{<\tau(s),\phi(e),\tau(t)>}}{\sqrt{d}}.\\
    K^i\left(s\right)=&K-Linear_{\tau(s)}^i\left(H^{(l-1)}[s]\right).
\end{align}

Linear projection $K-Linear:\mathbb{R}^d \rightarrow \mathbb{R}^{\frac{d}{h}}$ projects$\tau(s)$-type node $s$ into $Key$ vector $K^i\left(s\right)$, where $h$ is the number attention heads, $d$ is the overall vector dimension and $\frac{d}{h}$ is the vector dimension per head. 
\begin{equation*}
    Q^i\left(t\right)=Q-Linear_{\tau(t)}^i\left(H^{(l-1)}[t]\right).
\end{equation*} 
Similarly, $Q-Linear_{\tau(t)}^i$ is a linear projection projecting the target $node$ $t$ into the $i$-th Query vector $Q^i\left(t\right)$. $W^{Att}_{\phi(e)}\in\mathbb{R}^{\frac{d}{h}\times\frac{d}{h}}$ is a edge-based weight matrix for edge type $\phi(e)$.
\begin{equation*}
    Mes-head^i_{(s,e,t)}=M-Linear^i_{\tau(s)}\left(H^{(l-1)}[s]\right)W^{Mes}_{\phi(e)}
\end{equation*}
refers to the $i$-th message head. $M-Linear:\mathbb{R}^d \rightarrow \mathbb{R}^{\frac{d}{h}}$ is a linear projection. $W^{Mse}_{\phi(e)}\in\mathbb{R}^{\frac{d}{h}\times\frac{d}{h}}$ is a matrix incorporating edge dependency. $\mu\in\mathbb{R}^{\lvert A\rvert\times\lvert R\rvert\times\lvert A\rvert}$ is a adaptive scale tensor to attention.
% Then, we transmit $\widetilde{\Mat{H}}^{(l)}[t]$ through a Relu activation layer, make linear projection and make a residual connection in which way we can map the vector of the target node $t$ back to its specific type of attribute dimension, \ie,
Then the embedding of target node $t$ can be obtained by
\begin{equation}
    \Mat{H}^{(l)}[t]=\sigma \left(Linear_{\tau(t)}\widetilde{\Mat{H}}^{(l)}[t]\right)+\Mat{H}^{(l-1)}[t].
\end{equation}
where $\sigma$ is the relu function, $Linear_{\tau(t)}$ is a linear projection projecting target node $t$'s vector back to its specific type of attribute dimension.

In \emph{AHEAD}, we input one view of each node of one type into $HGT$ to be encoded separately, which means there are $K$ HGT models in total for target node $t$, where $K=\underset{\forall s\in N\left(t\right)\cup\{t\}}{\prod}k_{\tau(s)}$. Naturally, there are $K$ outputs of HGT, each of which is an embedding of one view of target node $t$.

We map attributes of each view into the same space before they are fed to HGT. For edge $(s,e,t)$, the process of the $i$-th view of $t$ and $j$-th view of $s$ fed to $HGT$ can be described as
\begin{equation}
    \begin{aligned}
    \Mat{H}^{(0)}[t_i]=&H-Linear_{\tau(t)}^i\left(\Mat{X}_{\tau(t)}^i\right),\\
    \Mat{H}^{(0)}[s_j]=&H-Linear_{\tau(t)}^i\left(\Mat{X}_{\tau(s)}^j\right),
    \end{aligned}
\end{equation}
where $H-Linear_{\tau(v)}^i:\mathbb{R}^{D_{\tau(v)}^i}\rightarrow\mathbb{R}^{h_1}$, $h_1$ is the hidden channels. Note that $H-Linear_{\tau(v)}^i$ is indexed by node $v$'s type, which s each type of node has a unique linear projection to compress its dimension of each view's attributes. $t_i$ denotes the $i$-th view of $t$ and $s_j$ denotes the $j$-th view of $s$.

\subsection{View-Level Attention Aggregator}
The embedding obtained by the $l$-layer $HGT$ of $t_i$ from $s_j$, denoted by $\Mat{H}^{(l,j)}[t_i]$, is obtained via
\begin{align*}
    \Mat{H}^{(l,j)}[t_i]&=\sigma\left(Linear_{\tau(t)}\tilde{\Mat{H}}^{(l,j)}[t_i]\right)+\Mat{H}^{(l-1,j)}[t_i],\\
    \tilde{\Mat{H}}^{(l,j)}[t_i]&=\underset{\forall s\in N(t)}{\sum}\boldsymbol{Att}_{edge}(s_j,e,t_i)\times\boldsymbol{Mes}(s_j,e,t_i).
\end{align*}
Intuitively, there are $K$ embeddings of $t$'s every view whose message comes from $t$'s neighbours' views. We decompress these $K$ embeddings into one uniform dimension space using a linear projection $Out-Linear:\mathbb{R}^{h_1}\rightarrow\mathbb{R}^{h_2}$, where $h_2$ is out channels.
\begin{equation}
    \Mat{Z}_{\tau(t)}^{(i,j)}=Out-Linear\left(\Mat{H}^{(l,j)}[t_i]\right),
\end{equation}
$\Mat{Z}_{\tau(t)}^{(i,j)}$ is a decompressed latent representation of $t_i$ obtained from $s_j$. For a view binary $(i,j)$, $j\in\left[1,k_{\tau(s)}\right]$, $i\in\left[1,k_{\tau(t)}\right]$, there are $K$ combinations. Each of them is denoted by $(k)$, where $k\in[1,K]$. Consequently, we can denote every $ \Mat{Z}_{\tau(t)}^{(i,j)}$ as $\Mat{Z}_{\tau(t)}^{(k)}$. Then we employ view-level attention to aggregate these  
$K$ decompressed embeddings. Because we hope our model can learn the importance of each view automatically so that views which are more important will be more involved in aggregating process. As a result, the final embedding of $t$ will illustrate $t_i$'s feature better. The latent representation of target node $t$ is aggregated based on view-level attention by:
\begin{equation}
    \Mat{Z}_{\tau(t)}=\underset{\forall k\in [1,K]}{\sum}\Vec{W}^{(k)}_{v}\times \Mat{Z}^{(k)}_{\tau(t)},
\end{equation}
where $\Vec{W}_{v}=(\Vec{W}_{v}^{(1)},\Vec{W}_{v}^{(2)},\cdots,\Vec{W}_{v}^{(K)})\in\mathbb{R}^{K}$ is a learnable weight vector representing view-level attention with 
\begin{equation*}    \Vec{W}_{v}^{(k)}=Softmax(\alpha)^{(k)},\quad k=1,2,\dots,K.
\end{equation*}
$\Vec{W}_{v}^{(k)}$, referring to importance of the $k$-th combination of views. If one coordinate of  $\Vec{W}_{v}$ is $0$, this combination of views will be ignored.
% $\Vec{W}}_{v}$ is regarded as a parameter required to be optimized and updated like other model parameters in the process of encoding.

\subsection{Heterogeneous Graph Decoder}
We intend to reconstruct the adjacency matrix, attribute matrix, and node-type matrix, so as to use reconstruction errors as indicators of anomaly.

\paragraph{\textbf{Structure Decoder.}}
The structure decoder intend to calculate the reconstructed adjacency matrix based on the output of encoder.
As introduced in preliminary, every edge type
has a unique adjacency matrix. Similarly, each edge type also has a unique reconstructed adjacency matrix. As a result, the structure decoder calculates the reconstructed adjacency matrix based on edge types, \ie,
\begin{equation}
    \tilde{\Mat{A}}_{\phi(e)}=sigmoid\left(\Mat{Z}_{\tau(s)}\cdot\Mat{Z}_{\tau(t)}^{T}\right),
\end{equation}
where $\tilde{\Mat{A}}_{\phi(e)}\in\mathbb{R}^{n_{\tau(s)}}\times\mathbb{R}^{n_{\tau(t)}}$ is reconstructed adjacency matrix of edge $<s,e,t>$.
The structure reconstruction loss $ L_s$ is calculated by
\begin{equation}
        L_s=\underset{\forall\phi(e)\in\mathbb{R}}{\sum}\left \|\Mat{A}_{\phi(e)}-\tilde{\Mat{A}}_{\phi(e)}\right \|_F
\end{equation}
\paragraph{\textbf{Attribute Decoder.}}
%\subsubsection{Attribute Decoder}
The attribute decoder intend to reconstruct attribute information of each view. Attributes information of $\tau(v)$-type nodes are reconstructed by a fully-connected layer which can be expressed as
\begin{equation}
    \tilde{\Mat{X}}_{\tau(v)}=f_{relu}(\Mat{Z}_{\tau(v)}W+B),
\end{equation}
where $W\in\mathbb{R}^{h_2\times D_{\tau(v)}}$, and $B\in\mathbb{R}^{n_{\tau(v)}\times D_{\tau(v)}}$ is a bias term.

The attribute reconstruction loss $ L_a$ is computed by
\begin{equation}
        L_a=\underset{\forall\tau(v)\in\mathbb{A}}{\sum}\left \|\Mat{X}_{\tau(v)}-\tilde{\Mat{X}}_{\tau(v)}\right \|_F.
\end{equation}
\paragraph{\textbf{Node-Type Decoder.}}
Node-type decoder makes our model more suitable for heterogeneous graphs because the heterogeneity of nodes is the main feature that distinguish heterogeneous graphs from homogeneous graphs.
We exploit one-hot encoder to encode node type information. %to make our framework illustrate heterogeneity of nodes. Because the heterogeneity of nodes is the main feature that distinguish heterogeneous graphs from homogeneous graphs, by allowing the model to identify the feature, our model can be more suitable for heterogeneous graphs.
For node $v\in\mathbb{V}$, this progress can be described as
\begin{equation}
    \Mat{T}[v]=One-hot Encoder\left(\tau(v)\right),
\end{equation}
where $\Mat{T}[v]\in\mathbb{R}^{\lvert\mathbb{A}\rvert}$ is an one-hot vector, in which only one coordinate corresponding to $\tau(v)$ is 1 and the others are 0. We can pile up vectors $\Mat{T}[v]$ of all nodes $v\in\mathbb{V}$ into a node-type matrix $\Mat{T}\in\mathbb{R}^{\lvert\mathbb{V}\rvert}\times\mathbb{R}^{\lvert\mathbb{A}\rvert}$. Node-type decoder is designed in order to reconstruct the node-type matrix $T$, . Another linear projection is exploited in node-type decoder. For node $v\in\mathbb{V}$, it projects the final latent representation $\Mat{Z_{\tau(v)}}$ into the reconstructed node-type vector
\begin{equation}
    \tilde{\Mat{T}}\left(v\right)=T-Linear_{\tau(v)}\left(Z_{\tau(v)}\right).
\end{equation}
Similarly, $T-Linear: \mathbb{R}^{h_2}\rightarrow\mathbb{R}^{\lvert\mathbb{A}\rvert}$ is indexed by the node $v$'s type $\tau(v)$, meaning that every type of nodes $v$ has a node-type linear projection to project its embedding into a node-type vector. In order to reflect the importance of each node type to the overall node type information, we design the third level attention: node-type attention in the progress of piling up reconstructed node-type vectors into one overall reconstructed node-type matrix $\tilde{\Mat{T}}$. $\tilde{\Mat{T}[v]}$ is one row of $\tilde{\Mat{T}}$ which represents node $v$'s node-type vector.
\begin{equation}
    \tilde{T}[v]=Softmax\left(\Vec{W}_T\cdot\tilde{T}\left(v\right)\right),
\end{equation}
where $\tilde{T}[v]\in\mathbb{R}^{\lvert\mathbb{A}\rvert}$, $\Vec{W}_T\in\mathbb{R}^{\lvert\mathbb{A}\rvert}$ is a weight vector representing node-type attention, and $\cdot$ is inner product. $\Vec{W}_T$ is a learnable weight vector similar to $\Vec{W}_v$ which we have introduced above.
The node-type reconstruction loss $L_T$ is computed by:
\begin{equation}
    L_T=\left \|T-\tilde{T}\right \|_F.
\end{equation}
%\subsection{Reconstruction Loss}
%Because topology, attribute, and node-type information may all become the key evidence for us to judge a node as an anomaly node. What's more, anomalous nodes can be anomalous from different perspectives. We compute the deviation degree from these aspects as loss, and quantify these three errors with the Frobenius norm, aiming to measure the quality of reconstruction numerically.
Consequently, the overall reconstruction loss $L$ is
\begin{equation}
    L=L_a+L_s+L_T.
\end{equation}
\begin{table*}[t]
\tabcolsep = 7.5pt
\begin{tabular}{lcccccccc}
\toprule[1.5pt]
\textbf{Methods} & \multicolumn{1}{l}{\textbf{Graph}} & \textbf{Autoencoder} & \textbf{Attention} & \textbf{Heterogeneity} & \multicolumn{1}{l}{\textbf{IMDB}} & \multicolumn{1}{l}{\textbf{CoAID}} & \multicolumn{1}{l}{\textbf{PolitiFact}} & \multicolumn{1}{l}{\textbf{GossipCop}} \\ \hline
\emph{MLPAE} &  &  &  &  & 0.9035 & 0.8588 & 0.4857 & 0.5158 \\
\emph{ONE} & \multicolumn{1}{l}{} &  &  &  & N/A & 0.3961 & 0.4495 & 0.4564 \\
\emph{GAAN} & \textbf{\checkmark} &  &  &  & 0.6220 & 0.2134 & 0.5704 & 0.5195 \\
\emph{GCNAE} & \textbf{\checkmark} & \textbf{\checkmark} &  &  & 0.9056 & 0.6224 & 0.4496 & 0.4706 \\
\emph{DOMINANT} & \textbf{\checkmark} & \textbf{\checkmark} &  &  & 0.8835 & 0.8297 & 0.5586 & 0.5001 \\
\emph{DONE} & \textbf{\checkmark} & \textbf{\checkmark} &  &  & 0.9063 & 0.6773 & 0.4776 & 0.5481 \\
\emph{ALARM} & \textbf{\checkmark} & \textbf{\checkmark} & \textbf{\checkmark} &  & 0.4605 & 0.2130 & 0.5472 & 0.4844 \\\hline
\textbf{AHEAD} & \textbf{\checkmark} & \multicolumn{1}{c}{\textbf{\checkmark}} & \textbf{\checkmark} & \multicolumn{1}{c}{\textbf{\checkmark}} & \textbf{0.9139} & \textbf{0.8890} & \textbf{0.6287} & \textbf{0.5716} \\ \bottomrule[1.5pt]
\end{tabular}
\caption{Characteristic and AUC scores of different anomaly detection methods on $IMDB$, $CoAID$, $PolitiFact$, $GossipCop$. Graph, attention, autoencoder and heterogeneity denotes whether the method involves $GNN$s, attention mechanism, autoencoders or heterogeneity procession. N/A means this method can't be implemented via \emph{PyGOD} \citep{pygod2022} on this dataset.}
\label{tab:results}
\end{table*}
\subsection{Anomaly Detection}
% The object of optimization process is to minimize the loss function which consists of three parts: structural loss, attribute loss, and node-type loss.
%The optimization process is accompanied by the minimization of the loss function composed by three parts:structural loss attribute loss and node-type loss:
With the convergence of $L$, \emph{AHEAD} can gradually grasp the latent anomalous information. 
After training process, we use the anomaly score to indicate the anomalous degree of a node $v$, which is shown as follow.
%After iterative optimization progresses, the indicator of nodes' anomalous degree: anomaly score of node $v_i$ in a relation $<v_i,e,u_j>$
\begin{equation}
    \begin{aligned}
    s(v)= & \lambda_1\left \|\Mat{A}_{\phi(e)}[v]-\tilde{\Mat{A}}_{\phi(e)}[v]\right\|_2\\
    & +\lambda_2\left \|\Mat{X}_{\tau(v)}[v]-\tilde{\Mat{X}}_{\tau(v)}[v]\right\|_2\\
    & +(1-\lambda_1-\lambda_2)\left \|\Mat{T}[v]-\tilde{\Mat{T}}[v]\right\|_2
    \end{aligned}
\end{equation}
where $\lambda_1,\lambda_2$ are hyper parameters for balancing the weight among the three terms. We can rank anomalies according to anomaly scores. Or from the perspective of probability, the following formula is the probability that node $v$ is anomalous
\begin{equation}
    p(v) ={\frac{s(v)}{\underset{\forall v\in\mathbb{V}}{\max}s(v)}}.
\end{equation}

\section{Experiments}
In this section, we empirically evaluate our model's 
performance on several real-world datasets.
\subsection{Experiment Setup}
\noindent \textbf{Anomaly generation.} We select some real-world heterogeneous datasets and inject some anomalies into them to obtain ground-truth abnormal nodes. The method of injecting anomalies which comes from \emph{DOMINANT} \citep{DOMINANT} is introduced in detail in the appendix. There are two kinds of anomalous nodes that can be constructed, attribute anomaly and structural anomaly. We inject both types of anomalies in every adopted dataset to ensure a comprehensive evaluation.
~\\
\noindent \textbf{Datasets.} We adopt heterogeneous information networks in these four datasets to empirically evaluate \textit{AHEAD} and baseline methods. These four datasets are $IMDB$ \cite{IMDB}, $PolitiFact$ \cite{shu2020fakenewsnet}, $Gossip Cop$ \cite{shu2020fakenewsnet}, $CoAID$ \citep{2020CoAID}.We present the statistical information of nodes, attributes, and anomalies about the four adopted datasets in the appendix.

\noindent \textbf{Compared Methods}
We select some anomaly detection methods to compare with \emph{AHEAD}.  $\boldsymbol{MLPAE}$ \citep{MLPAE}, $\boldsymbol{GCNAE}$ \citep{GCNAE}, 
$\boldsymbol{ALARM}$ \citep{ALARM}, 
$\boldsymbol{ONE}$ \citep{ONE}, 
$\boldsymbol{DOMINANT}$ \citep{DOMINANT}, 
$\boldsymbol{DONE}$ \citep{DONE}, $\boldsymbol{GAAN}$ \citep{GAAN}. Detailed descriptions of these methods are in the appendix.

\subsection{Experiment Results}
Table \ref{tab:results} presents characteristics of each anomaly detection method and shows the anomaly detection performance which is reflected by AUC scores on four real-world datasets. After observation and comparison, we draw the following conclusions.
\begin{itemize}
    \item The proposed \emph{AHEAD} model outperforms all other methods on $IMDB$, $CoAID$, $Politifact$ and $Gossipcop$. It is proved that the heterogeneity of graphs has a great interference with the anomaly detection task because the selected methods for comparison are used on homogeneous graphs. It is obvious that anomaly detection on heterogeneous graphs requires to deal with heterogeneous graphs in unique ways in order to better identify anomalies, which is also the advantage of our model \emph{AHEAD}.
    % \item Generally speaking, $GNN$-based methods perform excellently as a whole on every dataset. As a result, it's safe to conclude that using $GNN$ backbone can capture anomaly information on graphs better than $MLP$ backbone or $MF$ backbone.
    \item Despite the above defects, some methods based on autoencoder like $MLPAE$, $GCNAE$ $DOMINANT$ and $DONE$, perform very well on most datasets. We deduce that autoencoder can still play an important role because the heterogeneity of anomaly does not affect the reconstruction of normal instances.
    \item The performance of some methods like $ALARM$ and $GAAN$ fluctuates greatly. We propose that it is because the individual differences of each heterogeneous data set are too large, and these methods have poor generalization ability and can not learn each dataset well.
    \item Although the number and ratio of anomalies in each dataset are different and vary greatly, \emph{AHEAD} still shows strong performance and stability, which demonstrates that \emph{AHEAD} still has strong learning ability and generalization ability even in different application scenarios.
    \item The detection performance of all methods on $IMDB$ dataset is generally good and much better than that on other datasets. $IMDB$ dataset only contains injected anomalies while the other datasets still contain natural anomalies, Natural anomalies are various and hard to capture by a model, which affects the detection performance. It indicates that anomaly detection should start with the types of anomalies.
\end{itemize}

\begin{figure}
    \centering
    \includegraphics[width=\linewidth]{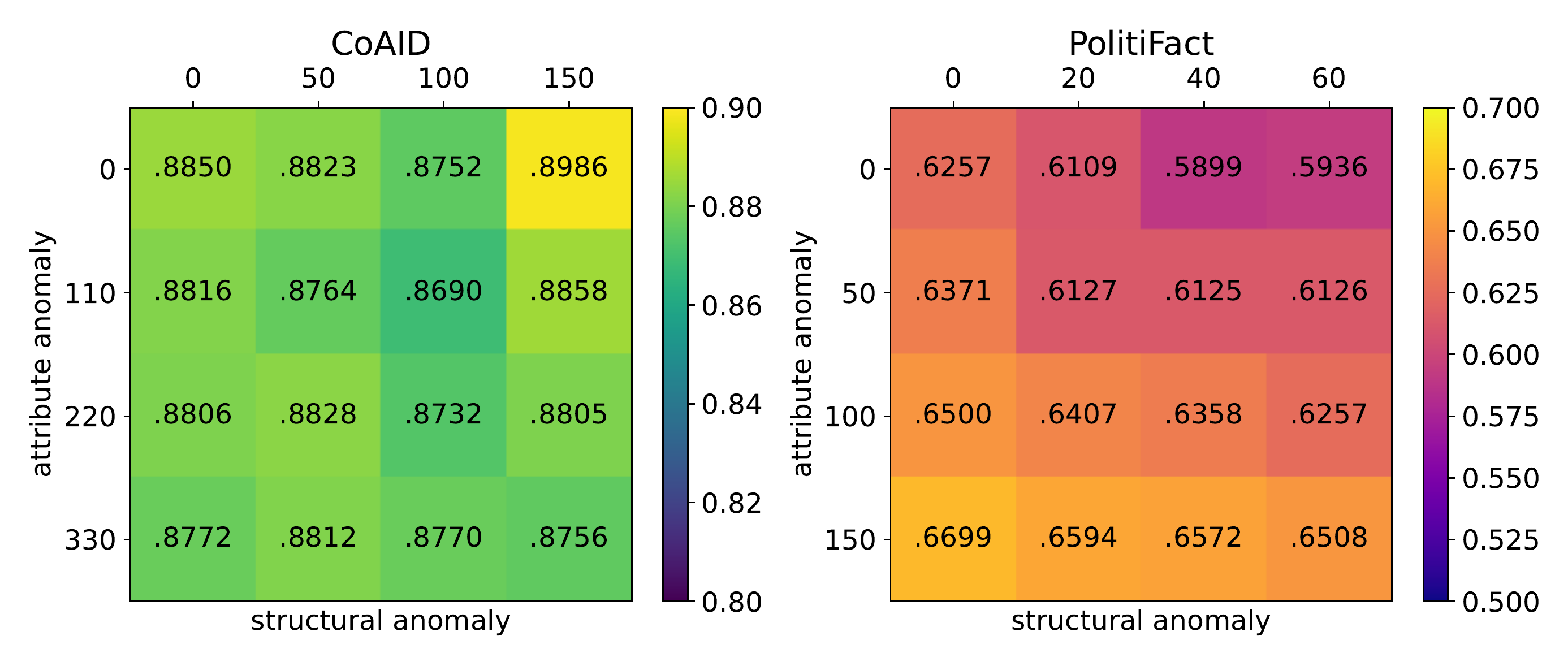}
    \caption{AUC scores of \emph{AHEAD} on $CoAID$ and $PolitiFact$ with different anomaly combinations. Attribute anomalies includes news and source. For example, 110 attribute anomalies consist of 100 news anomalies and 10 source anomalies.}
    \label{fig:robustness}
\end{figure}
\begin{figure}[t]
    \centering
    \includegraphics[width=0.9\linewidth]{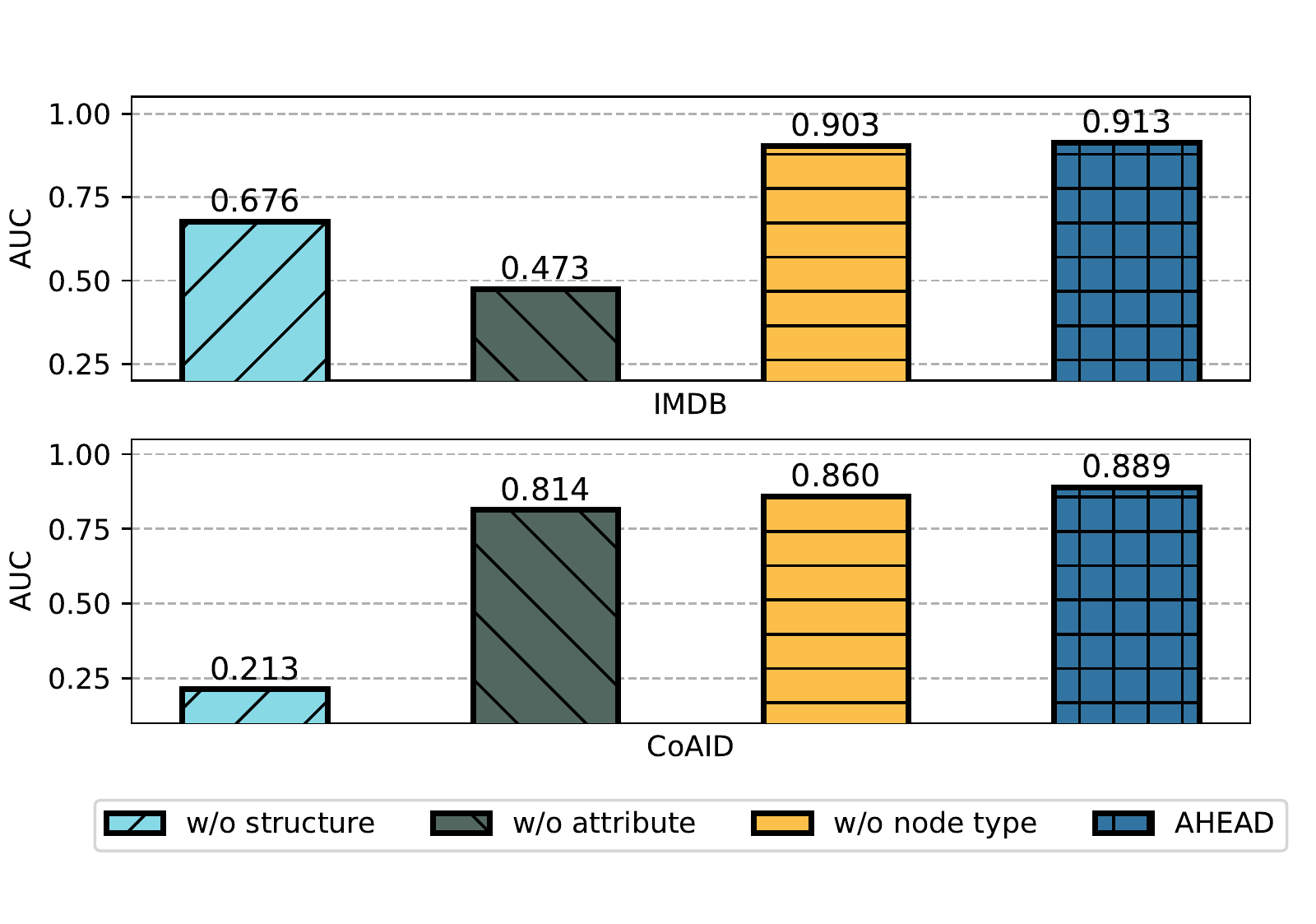}
    \caption{AUC scores of \emph{AHEAD} with different decoders.}
    \label{fig:abalation}
\end{figure}

\subsection{Effect of Decoders}
In order to explore the influence of each decoder on detection performance, we will delete one decoder in turn and observe the impact on the performance of the model. Fig. \ref{fig:abalation} shows the results.
We find that the performance of the model will decline in varying degrees when any kind of decoder is deleted. Only when the three decoders work at the same time, the performance of the model is the best. In addition, we find that each decoder plays a different role in different heterogeneous graphs, because in $IMDB$, AUC score of \emph{AHEAD} without attribute decoder decline significantly, while AUC score of \emph{AHEAD} without structure decoder 
drops a lot in $CoAID$. This is the reason for designing parameters $\lambda_1$, $\lambda_2$ when calculating the anomaly score to balance the three aspects.
\subsection{Robustness Test}
In practical applications, the number and type of anomaly can vary from time to time. In this subsection, we test the robustness of \emph{AHEAD} by changing the number and type of anomalies and observing how the performance of \emph{AHEAD} changes with it. Fig. \ref{fig:robustness} presents some experimental results of \emph{AHEAD} employed on $CoAID$ and $PolitiFact$ datasets with different anomalies in number and type. As is shown vividly in the Fig. \ref{fig:robustness}, although the number of anomalies of all types keeps rising, the performance of \emph{AHEAD} stays stable. AUC scores of \emph{AHEAD} on $CoAID$ fluctuate up and down by up to $2\%$, which is considered normal fluctuation. The fluctuation of detection performance on $PolitiFact$ dataset is slightly larger. This is what we anticipated in advance because $PolitiFact$ is small in size compared to $CoAID$ dataset. The anomaly radio will increase a lot if injecting anomalies. It may destroy the attribute and structure information of the dataset. Hence, The fluctuation is considered still within the normal range. Consequently, we deduce that \emph{AHEAD} is robust to the number and type of anomalies.

\subsection{Balance between structure \& attribute \& node-type}
\ 
\begin{figure}[]
    \centering
    \includegraphics[width=0.9\linewidth]{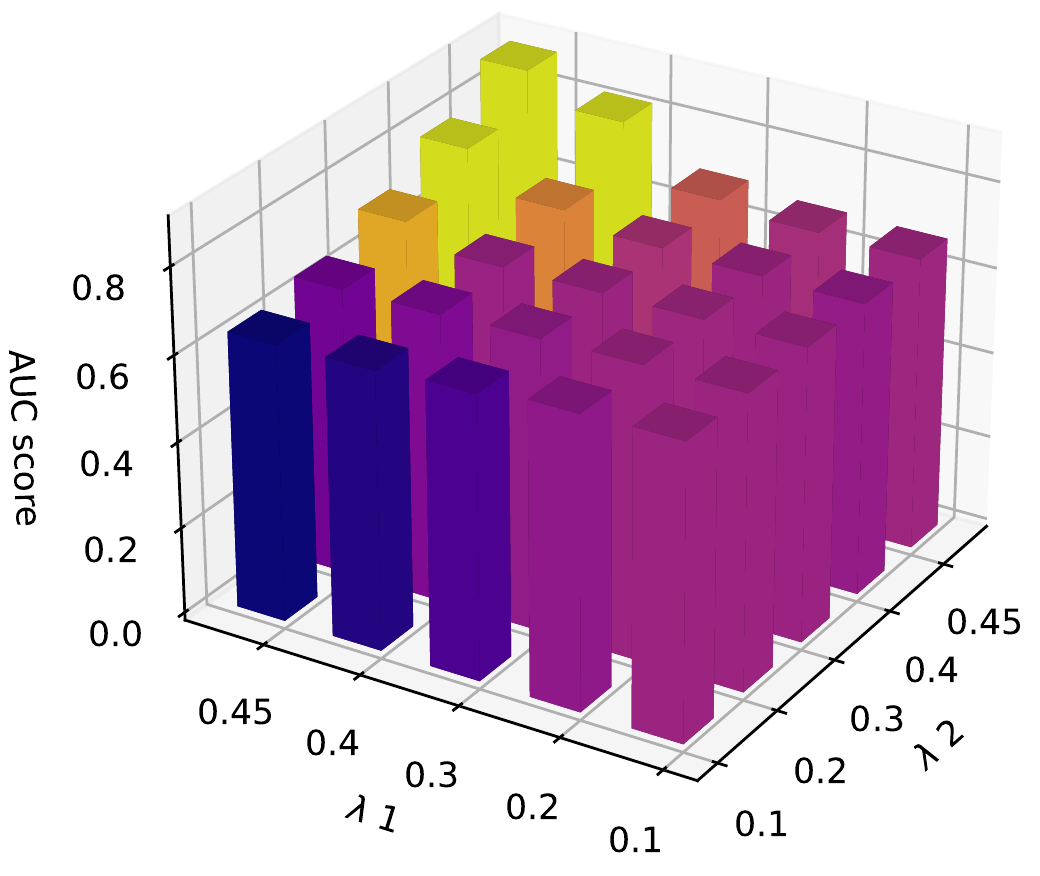}
    \caption{Influence of $\lambda_1$ and $\lambda_2$ on detection performance.}
    \label{fig:lambda}
\end{figure}

\noindent$\lambda_1$ and $\lambda_2$ are hyperparameters to balance three aspects of anomaly score: structure, attribute, and node-type. Here we explore how $\lambda_1$ and $\lambda_2$ influence the performance of \emph{AHEAD}, in other words, which aspect makes the greatest contribution to detecting anomalies. Experimentally, we select $IMDB$ as a representative dataset and change $\lambda_1$ and $\lambda_2$ separately from less to more and observe the performance of \emph{AHEAD}. There are some extreme situations. For example, when $\lambda_1$ equals zero, means that structure is not taken into consideration. These extreme situations have been discussed more deeply and comprehensively above, so they are not considered here. Fig.\ref{fig:lambda} presents experimental results.

After careful comparison, it's concluded that structure and attribute make contribution to anomaly detection more effectively than node-type. Generally speaking, the more structure and attribute is involved in anomaly score, the better performance is. However, it doesn't mean the node-type decoder is meaningless, which has been proved above. This conclusion inspires us that we should put more emphasis on structure and attribute when balancing these three aspects.

\section{Conclusion}
In this work, we propose \emph{AHEAD}, a triple attention based heterogeneous graph anomaly detection approach. \emph{AHEAD} can detect anomalies on heterogeneous graphs while former methods can't. Meanwhile, we design triple attention: edge-level attention, view-level attention, and node-type attention to assist detect anomalies from different perspectives. As a result, \emph{AHEAD} can distinguish anomalies caused by the heterogeneity of edges, nodes, and attribute characteristics. Experiments on four real-world datasets demonstrate the superiority and effectiveness of \emph{AHEAD} over other methods on performance and generalization ability. Furthermore, we prove the completeness and indispensability of our model's decoder by conducting an ablation study on three decoders. It's shown that \emph{AHEAD} without any decoder can't achieve the best performance.% All in all, for the first time, we examine anomalies from the perspective of heterogeneity, extend the anomaly detection task to the heterogeneity graphs, and propose a triple attention mechanism to better deal with the differences and implied connections between various heterogeneity, which has not been achieved by others yet.
% Meanwhile, our work also inherits the excellent achievements of previous work.
% In future work, we will be focused on
% \begin{itemize}
%     \item Reducing the time complexity and memory cost of the algorithm, and reducing the requirements of the algorithm for computing equipment. 
%     \item Improving the performance and efficiency of anomaly detection, and making the anomaly detection model more practical in applications. 
% \end{itemize} 
% \bibliographystyle{aaai23}
\bibliography{aaai23}
\appendix
\section{Compared Methods}
To prove the superiority of \emph{AHEAD} in anomaly detection performance, we select some anomaly detection methods to compare with \emph{AHEAD}. 
To the best of our knowledge, there is no existing anomaly detection method on heterogeneous graphs. As a result, we select competitive anomaly detection methods on homogeneous graphs as follows:
\begin{itemize}
    \item $\boldsymbol{MLPAE}$ \citep{MLPAE}: $MLPAE$ uses multi-layer perceptron autoencoder with nonlinear dimensionality reduction to do anomaly detection tasks.
    \item $\boldsymbol{GCNAE}$ \citep{GCNAE}: $GCNAE$ is an unsupervised learning method based on graphs. It proposes variational graph auto-encoder (VGAE) based on the variational auto-encoder (VAE). Embedding of the nodes in the graph can be obtained through a graph convolutional network encoder and a simple inner product decoder and can be used in downstream tasks like anomaly detection.
    \item $\boldsymbol{ALARM}$ \citep{ALARM}:$ALARM$ is a deep multi-view framework 
    
    proposing a multi-view encoder, an aggregator, and a structure \& attribute decoder. $ALARM$ cannot be employed on heterogeneous graphs so it lacks the processing of heterogeneity of multi-type nodes. That's the most outstanding difference between $ALARM$ and \emph{AHEAD}.
    \item $\boldsymbol{ONE}$ \citep{ONE}: $ONE$ proposes an unsupervised outlier aware network embedding algorithm for attributed networks. By this means, a robust network embedding can be obtained. By optimizing structure and attribute loss, downstream anomaly detection tasks can be performed.
    \item $\boldsymbol{DOMINANT}$ \citep{DOMINANT}: $DOMINANT$ is a deep unsupervised anomaly detector consisting of a shared graph convolutional encoder, a structure reconstruction decoder, and an attribute reconstruction decoder. Anomalies can be detected via the computing of structure \& anomaly score.
    \item $\boldsymbol{DONE}$ \citep{DONE}: $DONE$ includes an attribute autoencoder and a structure autoencoder. It minimizes an attribute proximity loss, an attribute homophily loss, a structure proximity loss, a structure homophily loss, and a combination loss to optimize the model. Then detect anomalies via computing the average anomaly score.
    \item $\boldsymbol{GAAN}$ \citep{GAAN}: $GAAN$ is a generative adversarial attribute network anomaly detection framework, including a generator module, an encoder module, and a discriminator module. Reconstruction error and real-sample identification confidence are employed to detect anomalous nodes. 
\end{itemize}
\section{Datasets}
\begin{table*}[t]
\tabcolsep = 10pt
\begin{tabular}{lccccc}
\toprule[1.5pt]
 & \textbf{Node} & \textbf{Attribute Dimension} & \textbf{Views} & \textbf{Anomaly Nodes} & \textbf{Anomaly Ratio} \\ 
\midrule[1pt]
\textbf{IMDB} & \textbf{M}: 4,278 \textbf{A}: 5,257 & \textbf{M}: 3,066 \textbf{A}: 3,066 & \textbf{M: 2 \textbf{A}: 3} & 240 & 2.50\% \\
\textbf{CoAID} & \textbf{N}: 5,457 \textbf{S}: 199 & \textbf{N}: 1536 \textbf{S}: 768 & \textbf{N}: 3 \textbf{S}: 2 & 962 & 17.01\% \\
\textbf{PolitiFact} & \textbf{N}: 1,054 \textbf{S}: 285 & \textbf{N}: 1,536 \textbf{S}: 768 & \textbf{N}: 3 \textbf{S}: 2 & 463 & 34.58\% \\
\textbf{GossipCop} & \textbf{N}: 22,140 \textbf{S}: 2,027 & \textbf{N}: 1,536 \textbf{S}: 768 & \textbf{N}: 3 \textbf{S}: 2 & 5,359 & 22.17\% \\ 
\bottomrule[1.5pt]
\end{tabular}
\caption{Statistics of four adopted datasets. $M$, $A$, $N$, $S$ stand for movie, actor, news, source respectively.}
\label{tab:datasets}
\end{table*}
We present the statistical information of nodes, attributes, and anomalies about the four adopted datasets in Table \ref{tab:datasets}. 
\begin{itemize}
    \item $IMDB$ \cite{IMDB} is a dataset about movies and television programs, including information such as cast, production crew, and plot summaries, which contains 4278 movies, 2081 directors, and 5257 actors. In our experiment we only select part of it, which consists of two types of nodes: movie and actor and edges connecting these nodes.
    \item $PolitiFact$ \cite{shu2020fakenewsnet} is a news websites where journalists select original statements to evaluate and then publish their findings on the PolitiFact website, where each statement receives a rating reflecting the authenticity. Two types of nodes are selected in our experiments: news and source in our experiments.
    \item $Gossip Cop$ \cite{shu2020fakenewsnet} is a website that fact-checked celebrity reporting. $Gossip Cop$ \cite{shu2020fakenewsnet} dataset includes news content, social context, spatiotemporal information and their ground truth labels. News and source nodes and their relations are selected in our experiments.
    \item $CoAID$ \citep{2020CoAID} is a diverse COVID-19 healthcare misinformation dataset, including fake news on websites and social platforms, along with users' social engagement about such news. It includes news, related user engagements, social platform posts about COVID-19, and ground truth labels. Similarly, news and source nodes are selected in our experiments.
\end{itemize}

\section{Data Pre-processing}
Since selected methods can only be exploited on homogeneous graphs, we preprocessed our heterogeneous graph datasets. In heterogeneous graphs, different types of nodes usually have different attribute dimensions. Consequently, we transform the feature dimension of all types of nodes into the same dimension. After transformation, the attribute dimensions of all kinds of nodes are the same, we can treat all nodes in the heterogeneous graph as the same kind of nodes, and slice their attribute matrices along the column direction to obtain the attribute matrices of all nodes. Then we can use the method mentioned in the previous section on these datasets.
As to views, since data do not contain natural multi-view attributes, we randomly split the attribute set into different views according to the given number of views. 
\section{Anomaly Generation}
The methods of generating attribute anomalies and structural anomalies in a given dataset are shown below:
\begin{itemize}
    \item attribute anomaly: We randomly select $n$ nodes as the attribute perturbation candidates. For each selected node $i$, we randomly select other $k$ nodes in the dataset and select the node $j$ whose attributes deviate the most from node i among the k nodes we selected \ie. $x_j=\underset{j\in selected\ k\ nodes}{argmax}\left \|x_i-x_j\right \|^2$. Then we substitute the attribute $x_i$ with $x_j$. Finally, node $i$ is labeled as an attribute anomaly.
    \item structural anomaly: We randomly select $m$ nodes from the network, and make them fully connected. All of the $m$ nodes in this group are labeled as anomalies. We repeat this process until $c$ clusters are generated. The total number of structural anomalies is $m\times c$.
\end{itemize}

\section{Implementation}
\begin{table}[t]
\tabcolsep = 16pt
\begin{tabular}{c|c}
\toprule[1.5pt]
\textbf{Hyperparameter} & \textbf{Value} \\ \hline
optimizer & \emph{Adam} \cite{adam} \\
learning rate & $5\times10^{-3}$ \\
hidden channels & 64 \\
out channels & 16 \\
model depth & 2 \\
weight decay & $10^{-5}$ \\
attention heads & 2 \\
max epochs & 100 \\ \bottomrule[1.5pt]
\end{tabular}
\caption{Hyperparameter settings of \emph{AHEAD}.}
\label{tab:implement}
\end{table}
$PyGOD$ \citep{pygod2022} is an anomaly detection on graphs python library. We use $PyGOD$ to implement some baselines. Our implementation is trained on a Intel(R) Xeon(R) Silver 4210R CPU $@$ 2.40GHz cpu.

\emph{MLPAE} \cite{MLPAE}, \emph{ONE} \cite{ONE}, \emph{GAAN} \cite{GAAN}, \emph{GCNAE} \cite{GCNAE} and \emph{DONE} \cite{DONE} are implemented via $PyGOD$. We use the default parameters when implementing the baselines with $PyGOD$ and train one hundred epochs each. \emph{ALARM} \cite{ALARM}, \emph{DOMINANT} \cite{DOMINANT} and \emph{AHEAD} are implemented by \emph{PyTorch} \cite{pytorch} and \emph{PyTorch Geometric} \cite{pyg}. In \emph{DOMINANT} \cite{DOMINANT} implementation, learning rate is set as $5\times10^{-3}$, hidden dimension is $64$, dropout rate is $0.3$, balance parameter $\lambda$ is $0.8$ and the model is trained $100$ epochs. In \emph{ALARM} \cite{ALARM}, the dimension of two hidden layers are $16$ and $8$ respectively. Learning rate is set as $5\times10^{-3}$, trade-off parameter $\lambda$ is set as $0.001$ and training epoch is $100$. The implementation details and hyperparameters of \emph{AHEAD} is listed in Table \ref{tab:implement}.
\section{Parameter Analysis}
In this part, we investigate the impact of the following two
parameters on model performance: (1) depth of encoder, (2) the learning rate of optimizer.
\subsubsection{Depth of Encoder}
~\\
\begin{figure}
    \centering
    \includegraphics[width=\linewidth]{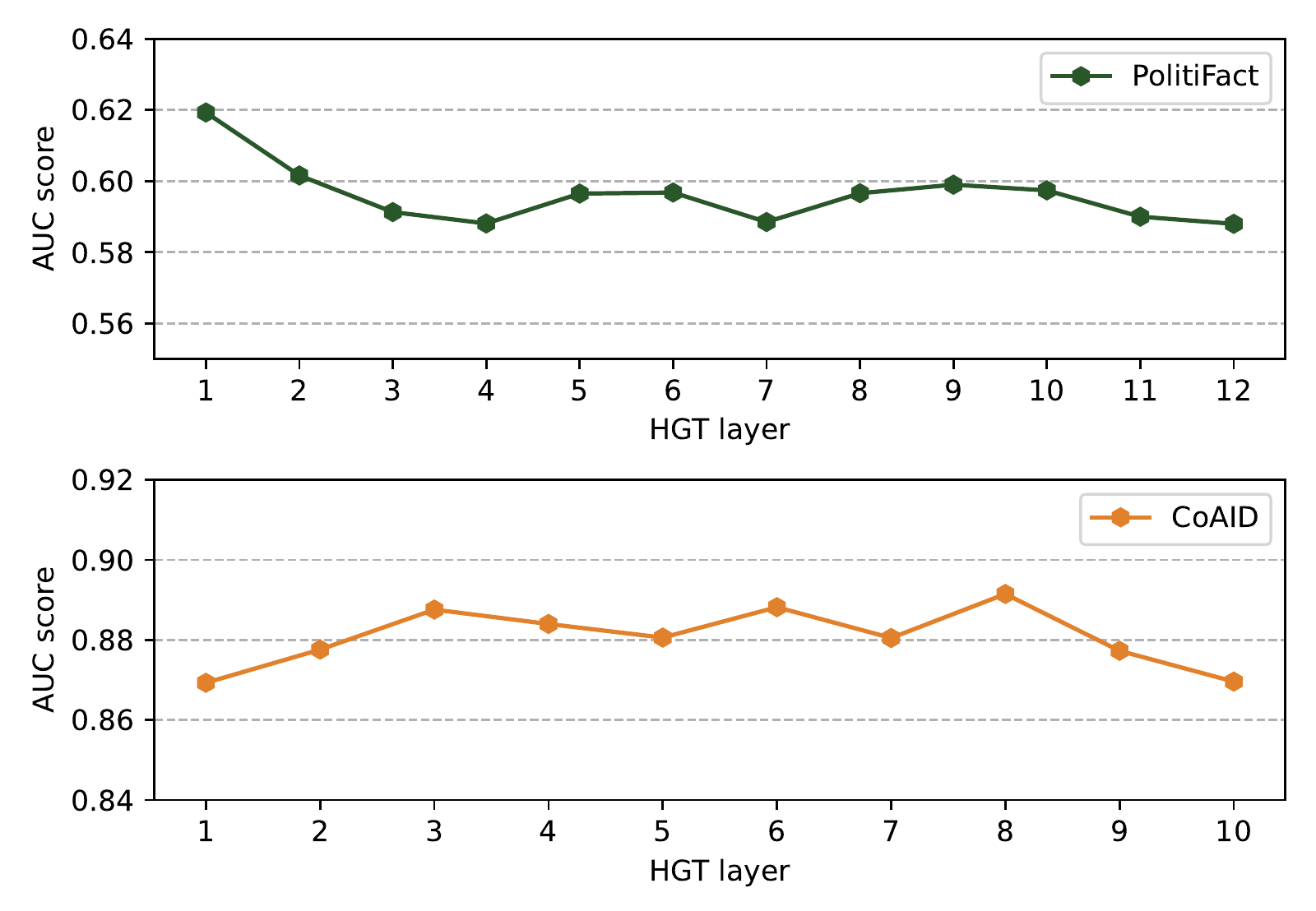}
    \caption{Influence of model depth on AUC scores on $PolitiFact$ and $CoAID$ datasets.}
    \label{fig:depth}
\end{figure}
Here we explore how the depth of \emph{AHEAD} influences the performance. We select $IMDB$ and $PolitiFact$ as representative datasets and adjust $HGT$ depth in the encoder and observe how AUC scores of \emph{AHEAD} fluctuate. Fig. \ref{fig:depth} presents experimental results. As can be observed easily, the best
performance on $PolitiFact$ dataset is achieved by a one-layer model while on $PolitiFact$ dataset is achieved by an eight-layer model. As the number of layers increases, the detection performance changes little at first, and then plummets a lot after a certain layer, indicating that overfitting may have occurred. This observation indicates that the model is not the deeper the better. The depth of the model needs to be carefully considered. Choosing an appropriate model layer can improve the performance of the model.

\subsubsection{Learning rate of optimizer}
\ 
\newline
\begin{figure}
    \centering
    \includegraphics[width=\linewidth]{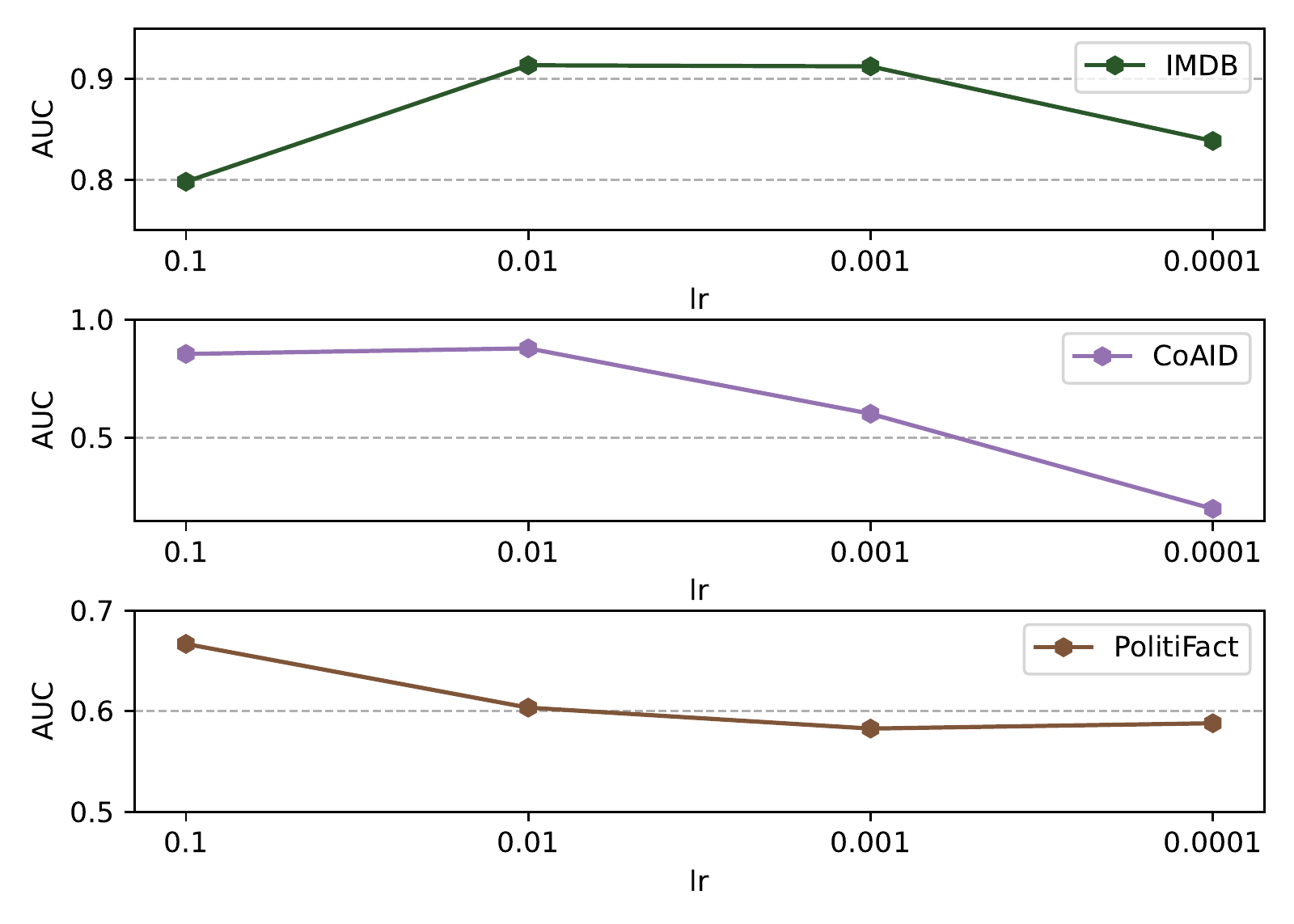}
    \caption{Influence of learning rate on detection performance on $IMDB$, $CoAID$ and $PolitiFact$ datasets.}
    \label{fig:lr}
\end{figure}
\noindent
Finally, we intend to explore the influence of learning rate on detection performance. In experiments, we adjust the learning rate of $Adam$ \cite{adam} optimizer and observe the detection performance on $IMDB$, $CoAID$, $PolitiFact$ datasets. We run one hundred epochs on each dataset. Fig. \ref{fig:lr} presents some experimental results. It can be easily observed learning rates influence detection performance significantly. Learning rates too high or too low will cause non-optimal solutions and bad performance. This indicates that in order to find the optimal solution faster and better, the learning rate needs to be carefully adjusted.

\end{document}